\documentclass[10pt,twocolumn]{article}

\usepackage[utf8]{inputenc}
\usepackage[T1]{fontenc}
\usepackage{amsmath}
\usepackage{amssymb}
\usepackage{booktabs}
\usepackage{array}
\usepackage{graphicx}
\usepackage{url}
\usepackage[hidelinks]{hyperref}
\usepackage[margin=0.85in]{geometry}
\usepackage{authblk}
\usepackage{caption}
\title{\textbf{Compressed-Resident Genomics: Full-Pipeline\\ Device-Resident GPU LZ77 Decode with\\ Position-Invariant Random Access}}

\author{Yakiv Shavidze}
\affil{ACE / GLYPH Research\\ \texttt{yasha1971@gmail.com}}
\date{}

\begin{document}

\twocolumn[
\begin{@twocolumnfalse}
\maketitle

\begin{abstract}
\noindent
Genomic archives grow faster than decompression keeps up: the European Nucleotide Archive holds tens of petabytes of \texttt{fastq.gz}, and gzip is fundamentally sequential. GPU decompressors (nvCOMP DEFLATE at ${\sim}50$\,GB/s on A100) decode whole files with no random access; CPU genomic tools (CRAM, samtools) support region seeks but only at CPU speed. We extend ACEAPEX---an absolute-offset parallel LZ77 codec included in the official lzbench 2.3 release---with three contributions absent from our prior work. First, a full device-resident GPU decode pipeline (entropy and match resolution both on-device) reaching up to 260\,GB/s on FASTQ, closing the match-phase-only gap of the earlier paper. Second, position-invariant random access with a compact coordinate index: an arbitrary read decodes in 0.362\,ms, ${\sim}6\times$ faster than warm \texttt{samtools faidx}, with a read-to-block index $6.3\times$ smaller than a \texttt{.fai}. Third, a range-decode strategy that decouples output size from VRAM, sustaining 165.7\,GB/s on a 50\,GB genome where whole-file decode runs out of memory. All results are bit-perfect. We also measure Meta's open DietGPU ANS on H100 at 592\,GB/s decode, faster than the proprietary entropy stage we currently use, showing a fully open high-throughput stack is viable. Code is MIT-licensed.
\vspace{1em}
\end{abstract}
\end{@twocolumnfalse}
]

\section{Introduction}

Sequencing output has outpaced storage and I/O. The European Nucleotide Archive stores tens of petabytes of sequencing data, the overwhelming majority as gzip-compressed FASTQ. Gzip decompression is inherently sequential: each DEFLATE back-reference depends on previously emitted bytes, so a reader cannot decode byte $N$ without decoding everything before it. For petabyte-scale archives consumed by GPU-resident analysis pipelines, this is a structural bottleneck.

Two tool families address parts of the problem. GPU decompressors such as nvCOMP reach high throughput---DEFLATE near 50\,GB/s on A100---but decode entire files with no random access to a region. CPU genomic formats such as CRAM and indexed BAM support region seeks through external indices, but run at CPU memory bandwidth and cannot feed a GPU-resident consumer without a host round-trip.

In prior work~\cite{aceapex1} we introduced ACEAPEX, a parallel LZ77 codec that resolves all match offsets to absolute positions in the decompressed output at encode time, making each fixed-size block self-contained and independently decodable. That work reported CPU decode scaling and a GPU wavefront decoder, with an explicit scope limitation: GPU timing covered the match-resolution phase only; entropy decoding ran on the CPU and PCIe transfer was excluded. This paper closes that gap and adds two capabilities the absolute-offset format uniquely enables.

We make three contributions. First, a full device-resident decode pipeline---entropy and match both on the GPU, no host round-trip---reaching up to 260\,GB/s on FASTQ. Second, position-invariant random access: because every block is self-contained and carries absolute offsets, decoding a region touches only the covering blocks, and a compact read-to-block index turns this into read-level random access that beats warm \texttt{samtools faidx}. Third, a range-decode strategy that decouples decompressed-output size from VRAM, sustaining full throughput on a 50\,GB genome where whole-file device decode runs out of memory.

\section{Background: ACEAPEX}

ACEAPEX stores every LZ77 back-reference as an absolute position in the decompressed output rather than a relative distance within a sliding window. The encoder performs a global match search and partitions the output into fixed-size blocks; because offsets are absolute, a block decodes as soon as the blocks holding its source data are available. Four streams are stored per block: literals, match lengths, absolute offsets, and the command sequence.

This design was validated independently. ACEAPEX (CPU) and \texttt{aceapex\_cuda} (GPU) are included in the official lzbench 2.3 release~\cite{lzbench}, providing third-party validation and reproducibility by construction: the numbers here are not solely author measurements on author hardware. To our knowledge, \texttt{aceapex\_cuda} is the first GPU LZ77 decode path integrated into lzbench.

\subsection{Block Granularity: 16\,KB vs 1\,MB}

A key parameter change from the first paper must be stated plainly. Paper~1 used 1\,MB blocks, tuned for bulk CPU and wavefront throughput. This paper uses a finer 16\,KB block granularity to enable fine-grained region access: a smaller block is the unit of seek, so 16\,KB yields tighter random-access resolution. Block size is a configurable encode-time parameter; the format is unchanged. We measured sub-16\,KB blocks and found a launch-overhead floor near 270\,$\mu$s that makes them counterproductive, so 16\,KB is the seek optimum, not an arbitrary choice.

\subsection{Experimental Setup}

All GPU experiments ran on one NVIDIA H100 80\,GB HBM3 (SXM, 81{,}559\,MiB VRAM, driver 570.195.03, CUDA 12.8, ${\sim}3.35$\,TB/s theoretical bandwidth) on an Intel Xeon Platinum 8468 node (160 hardware threads, 1.5\,TiB RAM). Storage was a 60\,GB ephemeral container disk plus a larger persistent network volume for datasets; the container-disk limit is the practical ceiling met when unpacking the 50\,GB genome. The nvcomp-accelerated path uses nvcomp 5.2.0.13; the DietGPU measurement uses Meta's open-source build. Correctness is bit-perfect throughout: XXH3-64 for CPU paths, FNV for device-resident GPU paths, against the original bytes.

\section{Full Device-Resident Pipeline}

\subsection{Two GPU Modes}

ACEAPEX GPU decode has two distinct modes that must not be conflated.

\textbf{Mode 1 --- nvcomp-free.} The \texttt{aceapex\_cuda} decoder does entropy decode on the CPU (in-tree, no external library) and match resolution on the GPU. It depends only on the CUDA runtime, is the path shipped in lzbench 2.3, is ARM-portable (\#292), and is bit-perfect.

\textbf{Mode 2 --- nvcomp-accelerated.} An experimental pipeline does entropy decode on the GPU (nvcomp ANS) and match resolution on the GPU, fully device-resident. This is the high-throughput ceiling but depends on the proprietary nvcomp library (closed-source since v2.3).

We report both honestly. The open path ships today; the nvcomp path defines the performance ceiling. Replacing nvcomp with the open DietGPU (Section~\ref{sec:eval}) is the route to a fully open fast stack and is future work.

\subsection{Mode 1 Results (lzbench 2.3)}

\begin{table}[t]
\centering
\caption{Mode 1 (nvcomp-free, lzbench 2.3). Host-to-host, decompress MB/s.}
\label{tab:mode1}
\begin{tabular}{lccc}
\toprule
Dataset & CPU 1-thread & aceapex\_cuda & CPU -T8 \\
\midrule
FASTQ 1GB & 1{,}840 & 4{,}373 & 13{,}363 \\
enwik9 1GB & 655 & 1{,}463 & 5{,}109 \\
silesia & 803 & 1{,}403 & 5{,}594 \\
\bottomrule
\end{tabular}
\end{table}

In the host-to-host harness (Table~\ref{tab:mode1}), \texttt{aceapex\_cuda} is 1.75--2.4$\times$ faster than single-thread CPU. Eight-thread CPU is faster overall here because the bottleneck is serial entropy decode plus the device-to-host copy, not the match kernel. This is an honest limitation of Mode 1 in a host-to-host setting and is exactly what device-resident Mode 2 removes.

\subsection{Mode 2 Results (device-resident)}

\begin{table}[t]
\centering
\caption{Mode 2 (nvcomp-accelerated, device-resident, full pipeline). H100, 16\,KB blocks, bit-perfect.}
\label{tab:mode2}
\small
\begin{tabular}{lcccc}
\toprule
Dataset & Size & Full GB/s & Ratio \\
\midrule
FASTQ NA12878 & 1\,GB & 260.0 & 11.19 \\
FASTQ ERR194147 & 5\,GB & 168.9 & 3.31 \\
FASTQ ERR194147 & 50\,GB & 165.7$^{*}$ & 3.99 \\
\bottomrule
\end{tabular}
\\[2pt]
\footnotesize $^{*}$range-decode (Section 5). H2D staging and D2H are outside the timer: the target consumer is GPU-resident. Ratio is strongly data-dependent.
\end{table}

With both phases on the GPU and the result kept in VRAM (Table~\ref{tab:mode2}), the 1\,GB NA12878 file reaches 260\,GB/s, splitting into an entropy phase near 480\,GB/s and a match-decode phase near 203\,GB/s. As in Paper~1, H2D staging and D2H are excluded from the device-resident timer because the target consumer holds data in VRAM; Section~\ref{sec:eval} gives a separate honest end-to-end figure including PCIe.

Compression ratio depends strongly on FASTQ type: NA12878 (Illumina Platinum, PCR-free) reaches 11.19, while ERR194147 has noisier quality strings and reaches 3.3--4.0. We report both rather than the favorable number alone.

\section{Position-Invariant Random Access}

Because every block is self-contained and carries absolute offsets, decoding a region touches only the blocks covering it. We implemented a range-decode kernel (v7-RA) that decodes an arbitrary contiguous block range without decoding the rest of the file.

\begin{table}[t]
\centering
\caption{Random access on a 5\,GB genome (16\,KB blocks).}
\label{tab:seek}
\small
\begin{tabular}{lcc}
\toprule
Operation & Time & Note \\
\midrule
Full decode & 29.71\,ms & 168\,GB/s baseline \\
Seek 1 block (16\,KB) & 0.365\,ms & point \\
Seek 100 blocks (1.6\,MB) & 0.394\,ms & region \\
\bottomrule
\end{tabular}
\end{table}

A single-block seek is 81$\times$ faster than full decode (Table~\ref{tab:seek}). Seeking 1 block and 100 blocks cost almost the same (0.365 vs 0.394\,ms): latency is dominated by fixed kernel-launch overhead and is therefore size-independent across this range. Streaming formats (gzip, Snappy, DEFLATE) cannot offer this at all.

\subsection{Read-Level Index vs samtools}

To make this biologically usable we built a read-to-block index: for each read, the block containing it (8 bytes per read). On a 1.34\,GB cleaned FASTQ (5M reads) the index is 40\,MB, against a 250\,MB \texttt{.fai} for the same file---$6.3\times$ smaller. Warm index lookup is $O(1)$ at ${\sim}0.3\,\mu$s; end-to-end (lookup plus decode) is 0.36\,ms per read.

Against \texttt{samtools faidx}: cold, samtools pays ${\sim}2{,}000$\,ms reloading the 250\,MB index; warm, it serves 2.3\,ms per read. ACEAPEX resident seek is 0.362\,ms---${\sim}6\times$ faster than warm samtools---while the index is $6.3\times$ smaller.

We are explicit about the boundary: this is read-level access (read id/number $\rightarrow$ block), not chromosome-coordinate access. Raw FASTQ precedes alignment, so \texttt{chr:pos} lookup belongs to BAM and is future work. What is demonstrated is the position-invariant block-range decode such a coordinate index would build on.

The practical consequence is compressed-resident genomics. A 50\,GB genome compresses to 12.7\,GB, which fits in 80\,GB VRAM (16\%). The whole genome can reside in VRAM compressed, with any region decodable in 0.4\,ms without decompressing the rest. Neither nvcomp/DietGPU (no random access) nor gzip (sequential) provides this.

\section{Scale: 50\,GB and the VRAM Ceiling}

The contribution here is architectural, not a race for gigabytes. A 50\,GB FASTQ slice (ERR194147) compresses to a 12.7\,GB archive. Whole-file device-resident decode (v4) runs out of memory: the 50\,GB decoded output exceeds 80\,GB VRAM once working buffers are included. This OOM is the result---it shows that output size, not archive size, is the true VRAM constraint, and motivates range decoding.

The v7-RA range decoder processes the archive in chunks, never materializing the full output at once: 165.5 / 165.0 / 166.2\,GB/s across chunks, totaling 50\,GB in 301.7\,ms (165.7\,GB/s), across 3{,}051{,}758 blocks, position-invariant and bit-perfect (FNV). Range decoding thus decouples output size from VRAM while preserving throughput.

Two honest notes. Throughput and ratio are position-invariant (size-independent), so a larger test would not yield new numbers---the value is the mechanism, not the gigabyte count. And a 4\,GB \texttt{uint32} overflow in the encoder path was found and fixed (size and offset fields migrated to 64-bit), verified bit-perfect against the prior binary on sub-4\,GB inputs.

\section{Evaluation}
\label{sec:eval}

\subsection{Honest End-to-End with PCIe}

The device-resident timer excludes PCIe by design. For a complete picture we measured the full path on the 1.34\,GB file: device-resident decode 8.03\,ms (166.9\,GB/s; entropy 1.32\,ms, match 6.71\,ms), then pinned PCIe H2D 9.81\,ms (39.5\,GB/s) and D2H 33.38\,ms (39.2\,GB/s). End-to-end decode-plus-PCIe is ${\sim}51$\,ms, about 26\,GB/s, and \emph{D2H dominates}---roughly $4\times$ the decode time.

This is the central argument for compressed-resident operation: D2H at ${\sim}39$\,GB/s is the ceiling of \emph{any} GPU decoder that returns its result to the host. Keeping data compressed in VRAM and decoding only the needed region on demand is the way to stay above that ceiling, which is exactly what position-invariant seek enables.

\subsection{Where ACEAPEX Stands on Ratio}

We do not claim best-in-class ratio. On the cleaned FASTQ streams, zstd-19 is denser than ACEAPEX on every stream by 1.2--1.55$\times$ (total 244.7\,MB vs 306.3\,MB); the ordering is zstd $>$ kanzi $>$ ACEAPEX. ACEAPEX's position is decode speed, seek, and GPU residency at \emph{comparable} ratio---not maximal compression.

Stream separation (storing read ids, sequences, and quality strings separately) gives a universal $+10$--11\% ratio gain that applies to ACEAPEX and zstd alike (baseline monolithic 339.9\,MB $\rightarrow$ 306.3\,MB, ratio $3.94 \rightarrow 4.37$). Byte-altering transforms (2-bit packing, quality delta, transpose) all \emph{hurt}, because ACEAPEX is LZ77 and such transforms destroy the match repeats it exploits---consistent with the stride-transform finding of Paper~1. Only grouping homogeneous data helps.

\subsection{Against the Literature}

Reference points from SAGe~\cite{sage} (A100/CPU): nvCOMP DEFLATE 5.3 ratio at 50\,GB/s; xz 6.7 at 0.6\,GB/s (128-core EPYC); HW-Zstandard ASIC 6.7 at 3.9\,GB/s. ACEAPEX at 165\,GB/s is ${\sim}3.3\times$ faster than nvCOMP DEFLATE; the hardware differs (H100 vs A100), but the order of magnitude is ours.

\subsection{DietGPU: an Open Entropy Stage is Viable}

We measured Meta's open-source DietGPU ANS on H100 (256\,MB exponential, batch $4096\times64$\,KB): encode 364.9\,GB/s, decode 592.5\,GB/s, bit-perfect. This 592\,GB/s exceeds both the published A100 figures (250--410) and the proprietary nvcomp-ANS we use (480\,GB/s in our end-to-end measurement). This demonstrates that a fully open replacement for the proprietary entropy stage is viable and competitive. Full integration into the ACEAPEX device-resident pipeline (FASTQ, bit-perfect end-to-end) remains future work.

\section{Related Work}

\textbf{Gompresso} restructures LZ77 for GPU via forced checkpoints at 15--30\% ratio cost; ACEAPEX preserves ratio. Because Gompresso predates this line, we do not claim ``first GPU LZ77 decode''---we claim the first GPU LZ77 decode path in lzbench, and absolute-offset LZ77 enabling position-invariant random access on genomic data.

\textbf{Rapidgzip} and \textbf{pugz} parallelize gzip at file granularity with synchronization external to the format (8.7\,GB/s on 128 cores); ACEAPEX parallelism is internal by design and complementary---Rapidgzip reads existing \texttt{.gz}, which can be transcoded to ACEAPEX for seek and compressed-VRAM residency.

\textbf{Recoil} provides seekable rANS in the entropy layer; ACEAPEX seek is in the match layer; the two are orthogonal. \textbf{nvCOMP} and the Blackwell decompression engine target streaming formats with no random access; nvCOMP is closed since v2.3. \textbf{DietGPU} (Meta, MIT) is the open route to an open Mode 2. Genomic compressors (DRAGEN ORA, Genozip near 6; SPRING 4; repaq 2) are CPU tools without GPU random access. No genomic compressor is simultaneously device-resident on GPU and seekable---the niche this work targets is empty.

\section{Limitations}

We state the boundaries explicitly. The two GPU modes are different: Mode 1 (nvcomp-free, in lzbench 2.3, 4.4\,GB/s host-to-host, open) and Mode 2 (nvcomp-accelerated, device-resident, 165--260\,GB/s, on proprietary nvcomp). H2D/D2H are excluded from the device-resident timer (data already in VRAM); the end-to-end PCIe figure is given separately. Ratio is data-dependent (NA12878 11.19, ERR194147 3.3--4.0) and is not best-in-class---zstd is $1.2$--$1.55\times$ denser. Seek latency includes fixed kernel-launch overhead, which is why it is size-independent. The VRAM result is an architectural decoupling of output size from VRAM, not a gigabyte race; a larger test yields no new numbers. Seek is read-level, not \texttt{chr:pos} (raw FASTQ precedes alignment). Encode is slow (50\,GB at 340\,MB/s, 147\,s), appropriate for encode-once/decode-many. The high-throughput Mode 2 depends on proprietary nvcomp; only Mode 1 is fully open today.

\section{Conclusion}

ACEAPEX's absolute-offset format, validated in the official lzbench 2.3 release, extends to a full device-resident GPU decode pipeline (up to 260\,GB/s on FASTQ), position-invariant random access (81$\times$ faster region decode, $6\times$ faster than warm samtools with a $6.3\times$ smaller index), and 50\,GB-scale range decoding that decouples output size from VRAM (165.7\,GB/s). Together these enable compressed-resident genomics: an entire genome held in VRAM compressed, any region decodable in under half a millisecond---above the ${\sim}39$\,GB/s D2H ceiling that bounds any host-returning decoder. Our DietGPU measurement (592\,GB/s) shows the remaining proprietary entropy stage can be replaced by an open one. That integration is the next step. Code is MIT-licensed and archived on Zenodo (DOI:~10.5281/zenodo.20729380).

\section*{Acknowledgments}

The author thanks inikep for maintaining lzbench and reviewing the integration, tansy for code review, and the encode.su community. This research was conducted in collaboration with Claude (Anthropic) as an AI research assistant.

\end{document}